# Inhomogeneity of the intrinsic magnetic field in superconducting YBa$_2$Cu$_3$O$_X$ compounds as revealed by rare-earth EPR-probe.

M.R. Gafurov[1,2], I.N. Kurkin[1] and S.P. Kurzin[1]

1. Laboratory of Magnetic Resonance, Physical Department of Kazan State University, 420008 Kazan, Russian Federation
2. I.Physikalisches Institut, Justus-Liebig Universität, D-35392 Giessen, Germany

**Abstract**

X-band electron paramagnetic resonance experiments on doped Er$^{3+}$ and Yb$^{3+}$ ions in YBa$_2$Cu$_3$O$_X$ (6 < X < 7) compounds with different oxygen contents in the wide temperature range (4 – 120 K) have been made. In the superconducting species at the temperatures significantly below T$_C$, the strong dependencies of the linewidth and resonance line position from the sweep direction of the applied magnetic field are revealed. The possible origins of the observed hysteresis are analyzed. Applicability of the presented EPR approach to extract information about the dynamics of the flux-line lattice and critical state parameters (critical current density, $J_C$, magnetic penetration depth, λ, and characteristic spatial scale of the inhomogeneity) is discussed.

## 1. Introduction

The magnetic behavior of high temperature superconductors (HTSC) is a fruitful area of scientific research. Particular attention is drawn to the inhomogeneity of the internal (local) magnetic fields revealed by different spectroscopy methods such as muon spin resonance spectroscopy (μSR) and nuclear magnetic resonance (NMR). The contributions from the vortex lattices, demagnetizing and pinning effects, the influence of the granularity are discussed; corresponding parameters are extracted; and different models of the distribution of the magnetic field are proposed in these investigations [1, 2]. One of the most examined substances is $YBa_2Cu_3O_x$ (YBaCuO, YBCO, 1-2-3) compound which is superconductive at $x > 6.35$ ($T_C \approx 92$ K for $x \approx 7.0$).

The applicability of electron paramagnetic resonance (EPR) for such kinds of experiments is restricted by the problem of EPR-silence in high $T_C$ cuprates [3]. There is no EPR response in pure $YBa_2Cu_3O_7$ compound, for example, while the nature of the own magnetic centers in the underdoped samples ($YBa_2Cu_3O_x$ with $x < 6.96$) or in the samples with worse quality is up to now under discussion. Moreover, the interactions of the own magnetic centers or incorporated spin probes (which lines, as a rule, are detected at $g \approx 2$) with charge carriers and between themselves make the EPR spectrum very complicated to interpret it.

Therefore, the *decoration of HTSC surfaces* with spin labels (1,1-diphenyl-2-picrylhydrazyl, DPPH, as usual) is used as a common technique for inhomogeneity studies with magnetic resonance methods. The lineshape, the position and narrow linewidth of DPPH itself are practically temperature independent. In the decoration experiments, however, they depend not only from the temperature but from other detection conditions, such as a frequency and a value of the modulation field, the sweep direction of the applied magnetic field, et al. This EPR technique in reference to HTSC, firstly presented by Rakvin et al. [4, 5], is being actively used and developed (see [6, 7, 8, 9, 10, 11], for instance).

In the paper [4], Rakvin et al. have ascribed DPPH linewidth broadening in $YBa_2Cu_3O_7$ just below $T_C$ to the inhomogeneity of the magnetic field produced by the Abrikosov vortex lattice. Penetration depth $\lambda_0 = 1600$ A was estimated. But in the next work [5] these authors have concluded that the inhomogeneity is mainly caused by the granularity of the samples and pinning on the boundaries of the superconducting grains is responsible for the irreversibility of the magnetic properties. The right value of $\lambda_0$ extracted in [4] was announced as an accidental coincidence. It had initiated the incessant discussion about the main contribution into the effects obtained and about the possibility to derive the corresponding parameters from the EPR experiments.

EPR investiagations of the own magnetic centers should help to solve this problem. Moreover, in this case the extracted parameters should represent the field distribution not only in the vicinity of the surface but inside the superconductor. Because of the restriction described above, we know only few papers in which the dependencies of the EPR lines of the own magnetic centers from the detection conditions were observed.

Badalyan and Baranov [12] have noticed the dependence of the EPR spectrum of $Gd^{3+}$ in $GdBa_2Cu_3O_X$ in parallel orientation (H ∥ C) at $T < T_C$ from the sweep direction as well as from the initial value of the applyied magnetic field at which the registration begins. The EPR line shift and broadening for the different detection conditions have been reported in $Rb_xC_{60}$ [13]. The small effect has been observed only due to the very narrow EPR line. The value of the critical current density $J_C$ was estimated. Hysteresis of $Cu^{2+}$ line was detected in $Tl_2Ba_2Ca_2Cu_3O_{10}$ (in the magnetic fields ≈ 8 κG at the frequency ≈ 24 GHz) by Nishida et al. [14]. The authors did not discuss this fact in details. Conduction EPR (CESR) of polycrystalline $MgB_2$ in the wide frequency range (3 – 225 GHz) [15] showed CESR line shift from the sweep direction below $T_C$ (40 K). The magnitude of the observed hysteresis decreased with frequency. The detailed discussion was not given as well. It is worth to notice again that in all papers cited in this paragraph, the EPR was observed at g ≈ 2 with all its inherent disadvantages mentioned above.

Consequently, we can summarize the EPR investigations of the inhomogeneous distribution of the magnetic field in HTSC as follows:

1. the experiments have been carried out mainly by using of the surface decoration technique;
2. the experiments have been carried out for the magnetic centers with g ≈ 2 where the friendly (useful) signal is often distorted by other interactions;
3. the origin of the EPR line broadening and of the magnetic hysteresis below $T_C$ is being hotly debated.

The first evidence of the dependence of the EPR spectra of the doped rare-earth ions with g ≠ 2 in superconducting species $YBa_2Cu_3O_X$ with different oxygen contents from the detection conditions (from the sweep direction of the applied magnetic field) is reported.

## 2. Experimental Details

X-band spectrometer IRES-1003 (9.25 – 9.48 GHz) was used in the wide temperature range (4 -120 K). The low microwave power (≈ $10^{-5}$ W) was applied in order to exclude the effects of saturation. The $YBa_2Cu_3O_X$ samples with 1% of rare-earth impurities were prepared by the standard solid-state reaction technique in the Laboratory of Magnetic Resonance of Kazan State University. The details of preparation are given in [16, 17, 18] where we have studied these species but not hysteresis phenomena.

$Er^{3+}$ and $Yb^{3+}$ dopants substitute $Y^{3+}$ giving the simple ($S_{eff}$ = 1/2) and sufficiently intensive EPR spectra in the easily achievable magnetic fields (g ≈ 3 – 3.4 for $Yb^{3+}$; g ≈ 4 – 8 for $Er^{3+}$ that corresponds to the resonance fields about 1000 - 2200 G). Therefore, their lines are not overlapped either with an unavoidable signal at g ≈ 2 (H ≈ 3400 G) from one side or with the low-field non-resonance signal of microwave absorption from other.

The exact value of the oxygen content $x$ was defined from the lattice parameter along the crystallographic *c*-axis [19] using X-ray diffraction. Purity checking of our samples by means of X-ray phase analysis does not reveal any impurity phases with the accuracy higher than 1%. The values of $T_C$ for different $x$ were determined from the temperature dependence of microwave absorption in a low magnetic field.

We have investigated five samples with different oxygen contents:

1) $YBa_2Cu_3O_{6.85}$ + 1% $Er^{3+}$ ($T_C$ = 85K);
2) $YBa_2Cu_3O_{6.85}$ + 1% $Yb^{3+}$ ($T_C$ = 85K);
3) $YBa_2Cu_3O_{6.67}$ + 1% $Yb^{3+}$ ($T_C$ = 65K);
4) $YBa_2Cu_3O_{6.45}$ + 1% $Yb^{3+}$ ($T_C$ = 40K);
5) $YBa_2Cu_3O_{6.12}$ + 1% $Er^{3+}$ (non-superconductive).

The properties of YBaCuO are strongly anisotropic but small skin depth (several tenth of μm), intensive low-field microwave absorption, and intensive noise generated by the vortex lattice hinder the EPR observation in single crystals and in the large grains (> 20 μm) of ceramics. In this work, therefore, the YBCO powders were milled, then mixed with paraffin or epoxy resin and placed in a glass tube in a strong magnetic field (≥ 15 kG) to prepare the quasi-single-crystal samples. The *c* – axes of the individual crystallites were predominantly oriented along the direction *C* of the aligning magnetic field after hardening of epoxy resin or paraffin. The cylindrical samples with the height of 5-20 mm and diameter of 3-4 mm were oriented either in axial or in radial direction.

The optical microscopy shows the variation of the grains sizes in the range from 1 to 5 μm; agglomerates up to 20 μm are also exist.

## 3. Results and discussion

### *3.1. EPR spectra*

As it is shown in Figure 1, the EPR parameters of the rare-earth probes such as a resonance field ($H_R$) and a linewidth ($\Delta H_{pp}$) depend on the sweep direction of the applied magnetic field in the superconducting samples. In the present paper, we name the sweep direction from the lower to the higher magnetic fields as "up" (usually applied sweep direction) and pass through the resonance from the higher to the lower magnetic fields as "down". The study of the hysteresis effect can be summarized as follows.

1. Hysteresis is observed in the superconducting species at the temperatures below $T_{irr}$ which is in its turn much below $T_C$. The values of the resonance fields of the "down" lines is lower or equal than those for the sweep "up", $H_R^{down} \leq H_R^{up}$. Below $T_{min}$, the EPR lines in parallel orientation broaden with temperature decreasing but the "down" line much slower than the "up" one, $\Delta H_{pp}^{down} \leq \Delta H_{pp}^{up}$. The values of $T_{irr}$ and $T_{min}$ are very close to each other but much lower $T_C$ (see Table 1).

**Table 1**

The values of $T_C$ (superconducting transition temperature), $T_{irr}$ (the temperature below which the hysteresis is revealed), $T_{min}$ (the temperature at which the linewidth of the rare-earth probe is minimal), and $\Delta H_{pp}^{min}$ (minimal EPR linewidth) of $Yb^{3+}$ и $Er^{3+}$ ions in $YBa_2Cu_3O_X$ samples with different oxygen contents $x$.

| X | 6.85 (+Er) | 6.85 (+Yb) | 6.67 (+Yb) | 6.45 (+Yb) |
|---|---|---|---|---|
| $T_C$, K | 85 | 85 | 65 | 40 |
| $T_{irr}$, K | 35 | 60 | 25 | 20 |
| $T_{min}$, K | 25 | 55 | 30 | 25 |
| $\Delta H_{pp}^{min}$, G | 230 | 95 | 75 | 75 |

2. Hysteresis is observed in parallel orientation of the applied magnetic field, $H \parallel C$. There is no hysteresis in perpendicular orientation, $H \perp C$.
3. The magnitude of hysteresis *does not depend* from the following conditions:
   a. the sweep range (500 - 6400 G) and the sweep time (10 - 600 c);
   b. the initial value of the applied magnetic field (100 - 7000 G) in which heating or cooling of the sample occurs (field-cooling conditions);
   c. the temperature sweep direction (heating or cooling);
   d. the magnitude and the frequency of the magnetic field modulation (0-33 G at the frequencies 100 and 500 kHz, i.e. the hysteresis is observed even without modulation).
4. Hysteresis of the unavoidable spectrum with $g \approx 2$ ($H \approx 3400$ G, "Cu-like" signal) is not observed.

## 3.2. The resonance field ( g - factor)

The temperature and sweep dependencies of the resonance field of $Yb^{3+}$ ions in the superconductive samples with $x = 6.85$ and 6.67 are presented in Figure 2. The low temperature experimental values of perpendicular ($g_\perp$) and parallel ($g_\parallel$) components of g–factor, which correspond to the mean values of the resonance fields, $H_{R0} = \frac{H_R^{up} + H_R^{down}}{2}$, are listed in Table 2. These data are in an excellent agreement with the calculated values extracted by using the crystalline electric field parameters as determined by inelastic neutron scattering [20, 21]. The calculated data for the lowest doublets of $^4I_{15/2}$ and $^2F_{7/2}$ ground terms of $Er^{3+}$ and $Yb^{3+}$ ions, correspondingly, are listed in Table 3.

**Table 2**

The experimental values of parallel ($g_\parallel$) and perpendicular ($g_\perp$) components of g-factors of $Er^{3+}$ and $Yb^{3+}$ ions in $YBa_2Cu_3O_X$ at T = 30 K. The values of $g_\parallel$ are correspond to the mean values of the resonance field $H_{R0} = \frac{H_R^{up} + H_R^{down}}{2}$.

| X | 6.85 (+Er) | 6.12(+Er) | 6.85(+Yb) | 6.67(+Yb) | 6.45(+Yb) |
|---|---|---|---|---|---|
| $g_\parallel$ | 4.3(1) | 4.9(1) | 3.07 | 3.11 | 3.18 |
| $g_\perp$ | 7.6(1) | 7.15(10) | 3.52 | 3.49 | 3.48 |

**Table 3**

The calculated components of g-factors of $Yb^{3+}$ and $Er^{3+}$ in $YBa_2Cu_3O_X$ extracted using the crystalline electric field parameters as determined by inelastic neutron scattering [20].

| X | 6.98 (+Er) | 6.09 (+Er) | 6.91 (+Yb) | 6.78 (+Yb) | 6.45 (+Yb) |
|---|---|---|---|---|---|
| $g_\parallel$ | 4.28 | 4.89 | 3.107 | 3.134 | 3.243 |
| $g_\perp = (g_{xx}+ g_{yy})/ 2$ | 7.69 | 7.45 | 3.556 | 3.557 | 3.512 |

The approaching of the EPR lines in parallel and perpendicular orientations to each other at $T > T_{irr}$ could be ascribed to the temperature dependence of the lattice parameters of YBaCuO [22] as well as to the demagnetizing effects. The detailed discussion of this is not in the scope of the present paper.

The hysteresis at $T < T_{irr}$ can be likely caused by the flux-line pinning. For the sweep "up", due to the existing pinning centers, internal magnetic field in the sample is lower than the applied one. Therefore, the resonance condition is reached in the higher magnetic fields, $H_R^{up} > H_{R0}$ and the detected EPR line is shifted to the higher magnetic fields. For the sweep "down", conversely, the detected EPR line is shifted to the lower magnetic fields, $H_R^{down} < H_{R0}$.

The symmetry of the „up" and „down" shifts (cf. Fig 2) allow us to use a Bean model [23]. The gradient of the magnetic field is connected with the critical current density $J_C$

$$\frac{d\mathrm{H}}{d\mathbf{x}} = \frac{4\pi}{c} J_C \quad (1)$$

and can be approximately expressed as

$$\frac{d\mathrm{H}}{d\mathbf{x}} \approx \frac{\tfrac{1}{2}\left(H_R^{up} - H_R^{down}\right)}{\tfrac{1}{2}l} \equiv \frac{H_R^{up} - H_R^{down}}{l} \equiv \frac{\Delta H_R}{l}, \quad (2)$$

where $l$ is a characteristic spatial length scale of the pinning structure (inhomogeneity).

From equations (1) and (2) follow

$$\Delta H_R \propto J_C \quad (3)$$

and

$$J_C \left\langle \tfrac{A}{cm^2} \right\rangle = \frac{10 \cdot \Delta H_R \langle G \rangle}{4\pi l \langle cm \rangle}. \quad (4)$$

We have found that the temperature dependencies of $\Delta H_R \equiv H_R^{up} - H_R^{down}$ for the samples with different oxygen contents can be good described as (see Figure 3)

$$\Delta H_R = \Delta H_{R0} \cdot \exp\left[-3\left(\frac{T}{T_{irr}}\right)^2\right], \quad (5)$$

with the values of $T_{irr}$ listed in Table 1.

Corresponding to equation (5) expression

$$J_C = J_{C0} \cdot \exp\left[-3\left(\frac{T}{T^*}\right)^2\right] \quad (6)$$

was introduced in [24] for the temperature dependence of the critical current density in the line disorder pinning regime of HTSC. As it was shown in [25], equation (6) describes very well $J_C(T)$ behavior in $YBa_2Cu_3O_7 - Y_2BaCuO_5$ composites at the intermediate temperatures (40 – 75 K) in the magnetic fields lower than 1 T. In our experiments, this description is applicable for the different oxygen contents in the wider temperature range (5 K ≤ T ≤ $T_{irr}$).

Equation (4) establishes the relation between the critical current density $J_C$ and characteristic spatial scale of the inhomogeneity $l$. This relation at T ≈ 5 K is shown in Figure 4.

Davidov et al. [7] emphasize that in ceramic superconductors a variety of possible mechanisms for flux distribution exists.

1. The Abrikosov lattice is characterized by a typical intervortex distance of 0.1 μm.
2. The random grain structure is responsible for a flux distribution with $l$ of the order of the grain size (1 – 20 μm).
3. Intergrain flux distribution might provide larger length scales (tenth of micrometers).
4. The flux penetration via the sample edges or an inhomogeneous demagnetizing field lead to a macroscopic flux distribution on the sample size scale (a few millimeters).

The presented in this paper approach gives a principal opportunity to define $l$ and corresponding flux-pinning mechanism by mean of EPR and critical current density measurements. We do not measure $J_C$ by other methods in this work. Nevertheless, we can make a reasonable evaluation. Literature values of $J_{C0}$ are vary in a wide range from sample to sample in a low-field region: from $10^4 – 10^5$ A/cm$^2$ for X ≈ 6.5 in crystals up to ≈ $10^7$ - $10^8$ A/cm$^2$ in films [25, 26, 27]. Hence (cf. Figure 4), it is most likely that either *flux-line lattice* or a *grain structure* of our samples is responsible for the effects observed.

The influence of the flux-line lattice (FLL) on the EPR line shift is usually taken into account as [28, 4]

$$H_R^{up} - H_{R0} \equiv H_{R0} - H_R^{down} \equiv \frac{H_R^{up} - H_R^{down}}{2} = 0.0367 \frac{\Phi_0}{\lambda^2}, \quad (7)$$

where $\Phi_0 = h/2e = 2.07 \cdot 10^{-7}$ G·cm$^2$ is a magnetic flux quantum and λ(cm) is a magnetic penetration depth. From this equation follows that values of λ at T = 5 K ($\lambda_0$) are equal to (0.14 ± 0.02) μm for the samples with $x$ = 6.85 and 6.67; $\lambda_0$ = (0.30 ± 0.03) μm for $x$ = 6.45. The extracted values are discussed below, in Sec. 3.3, together with the data derived from the temperature dependence of the EPR linewidth.

Using equation (3) we can understand the absence of hysteresis in orientation H⊥C. Usually, the value of $J_C$ in perpendicular orientation ten times smaller than in a parallel one [25, 26]. Therefore, the magnitude of $\Delta H_R$ for H⊥C could not exceed 9 G even in the high-doped samples at T = 5 K. Taking into account a large linewidth (more than 100 G) and a measurement error (≈ 5 G), the hysteresis might be not detected.

One of the simplest explanations of the absence of hysteresis for the «Cu-like» signal ($H_R$ ≈ 3400 G vs. 1000-2200 G for the rare-earth ions) in our measurements is that the position of the irreversibility line is dependent from H. It qualitatively agrees with the results of paper [29] in which it is also shown that the position of the irreversibility line in H-T diagram strongly depends from the measurement method applied.

The experiments at the extremely low temperatures and at the low frequencies (weak magnetic fields) could prove or cut out these hypotheses.

### *3.3. EPR Linewidth*

Non-linear EPR line broadening at T > $T_{min}$ caused by the Orbach-Aminov (for $Er^{3+}$ ions) and the Raman (for $Yb^{3+}$ ions) processes of spin-lattice relaxation was discussed in [17, 18, 30] in details and is not presented in this paper. The values of $\Delta H_{pp}^{min}$ do not correlate with the values of oxygen content, $x$ (see Table 1), and additional experiments at the different frequencies in order to extract the homogeneous and the inhomogeneous parts of the line broadening are necessary.

With temperature decreasing below $T_{min}$, the line is broadening and this broadening depends on the sweep direction (see Figure 5). Some possible mechanisms of this are discussed below.

In paper [16] this temperature behavior (the line narrowing with temperature increasing from 4 K up to $T_{min}$) for the $Er^{3+}$ and $Yb^{3+}$ doped samples with $x = 6.85$ was explained by means of *thermoactivated motion*. It was suggested that the *fluctuating local magnetic field* on the rare-earth probes is due to the magnetic moments of $Cu^{2+}$ in the planes $CuO_2$. This field with the value $\sqrt{(h_0)^2}$ ≈ 160 G is directed along the *C*-axis and does not depend from the sort of the EPR probe. The frequency of the fluctuations can be described as

$$W = W_0 \exp(\tfrac{-U}{T}), \tag{8}$$

where $W_0$=(3.5 ± 0.7)·$10^{10}$ $s^{-1}$ and $U$ = (25 ± 3) K.

The EPR in [16] was detected only „up". The present results force us to revise the former conclusion because it is not so easy to assume that the local magnetic field from $Cu^{2+}$ ions depends from the sweep direction.

A review of possible reasons of the fluctuating magnetic fields in cuprates is given in [31] as well as a new theory is proposed. The calculations based on the relaxation measurements by NMR, EPR, and μSR by using these different theories give very similar to [16] results: $\sqrt{(h_0)^2} = (150 - 250)$ G and $U = (19 - 30)$ K. Nevertheless, it is not clear again how these local fields can depend from the sweep direction.

The inhomogeneity caused by Abrikosov's or Josephson's FLL leads to the EPR line broadening (see [28, 11], for example)

$$\Delta H_{pp} - \Delta H_{pp}^{\min} = 0.120 \frac{\Phi_0}{\lambda^2}. \tag{9}$$

There are some problems to ascribe our data either for the sweep "up" or "down" only to the influence of the FLL.

*(a)* From equations (7) and (9) follows

$$\Delta H_{pp} - \Delta H_{pp}^{\min} \approx 3.3\left(H_R^{up} - H_{R0}\right), \tag{10}$$

i.e. resonance line shift should be approximately three times less than the line broadening. Our results do not correlate with this formal relationship. The proportion factor varies from sample to sample.

*(b)* The temperature dependence of λ is usually taken into account as

$$\lambda(T) = \lambda_0 \left(1 - \frac{T}{T_C}\right)^{-1/2} \tag{11}$$

for the *d*-wave superconductivity or

$$\lambda(T) = \lambda_0 \left(1 - \left(\frac{T}{T_C}\right)^4\right)^{-1/2} \tag{12}$$

for the *s*-wave pairing.

According to equations (11) and (12), the influence of the FLL should manifest right below $T_C$ as revealed in the most experiments mentioned above. In papers [8, 11] as well as in our experiments, in contrary, the hysteresis and (or) the line broadening are observed below $T_{irr} < T_C$. The formal approximations of the linewidth by using equation (9) and (11) are presented in Figure 5. (The approximations by using equation (12) give even poorer results.) To explain the line broadening by the influence of the flux-line lattice, the authors of [11] were forced to replace $T_C$ in equation (11) with $T_{\min}$.

*(c)* The extracted by different methods literature value of $\lambda_0 \equiv \lambda_{ab0}$ in YBa$_2$Cu$_3$O$_7$ for H ∥ C is in the range (0.12-0.16) μm and increases with *x* reducing (see [32], for example).

Our formally calculated values of $\lambda_0$ by using equation (9) practically do not depend from *x*: $\lambda_0^{up}$ = (0.10 - 0.11) μm for the sweep "up" and $\lambda_0^{down}$ = (0.15 - 0.18) μm for the sweep "down" for all investigated samples. The authors of [33] have concluded that the fluctuations of the FLL could modify λ and an effective value of the penetration depth instead of the real one should be used to describe their NMR experiments in $Tl_2Ba_2Ca_2Cu_3O_{10}$. Nevertheless, it is still very hard to explain the independency of λ from the oxygen content and the discrepancy with the values of the penetration depths extracted from the line shift (see Sec. 3.2). Moreover, it is difficult to understand the sweep dependence in the framework of the presented approach. Therefore, we can conclude that the complicated temperature and sweep behavior of the linewidth observed can not be ascribed only to the influence of the flux-line lattice.

## Conclusion

We have managed to observe the EPR hysteresis in spite of the broad linewidth on the rare-earth paramagnetic probes $Er^{3+}$ and $Yb^{3+}$ inside of the superconducting samples. Their lines are not overlapped either with the unavoidable signal at g ≈ 2 from one side or with the low-field non-resonance signal of microwave absorption from other. The components of g-tensor can be good calculated by means of crystal field theory. Neither electron Knight shift of the resonance field nor Korringa slope of the linewidth is detected. It means that the interactions of these ions with charge carriers and between themselves are negligible small. From this point of view, EPR of Er and Yb ions is very suitable for the investigation of the distribution of the internal magnetic fields in HTSC.

For the thorough understanding of the experiment, measurements on the samples with different (calibrated) granular sizes not only by the EPR technique at the different frequencies but also by other methods are necessary.

## Acknowledgments

We are thankful to I. Kh. Salikhov, M.V. Eremin, V.A. Ivanshin and Yu.I. Talanov for their help with the preparation of the paper. This work was supported in part by CDRF grant REC-007 and scientific Fund of Republic of Tatarstan (NIOKR RT), Russia.

Figure 1

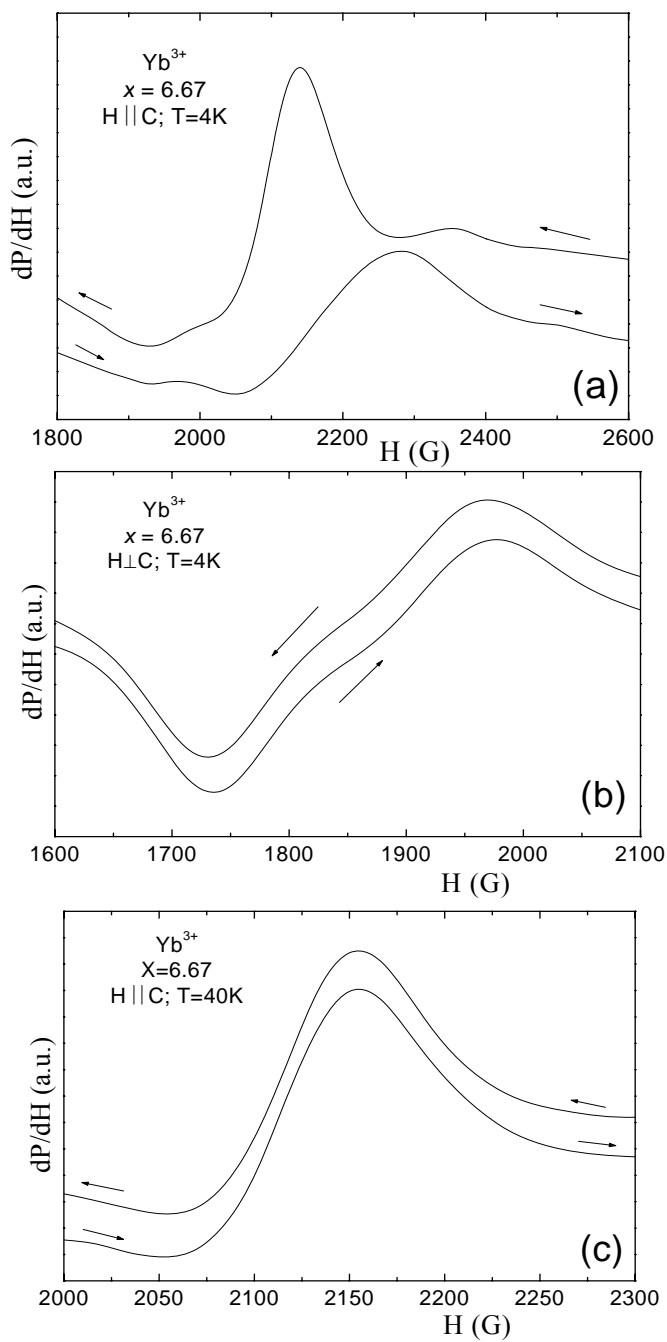

Fig. 1. The EPR spectra of $Yb^{3+}$ in $Yb_{0.01}Y_{0.99}Ba_2Cu_3O_{6.67}$ in parallel and perpendicular orientations at T = 4 K and 40 K. Arrows ($\rightarrow$ and $\leftarrow$) here and elsewhere show the sweep direction of the applied magnetic field ("up" and "down", correspondingly).

Figure 2

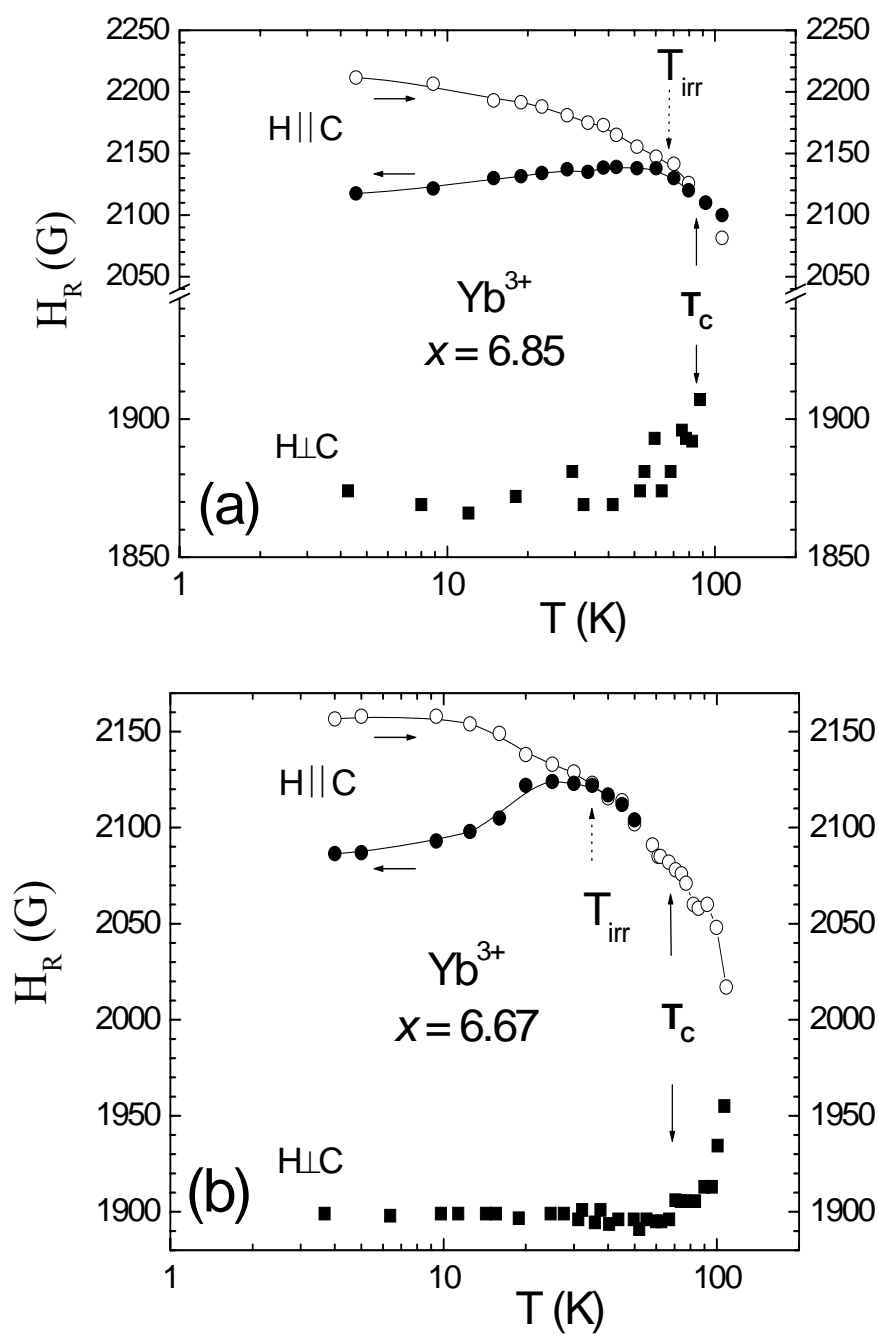

Fig. 2. Temperature and sweep dependencies of the resonance fields of $Yb^{3+}$ in $Yb_{0.01}Y_{0.99}Ba_2Cu_3O_{6.85}$ (a) $Yb_{0.01}Y_{0.99}Ba_2Cu_3O_{6.85}$ (b) in parallel and perpendicular orientations.

Figure 3

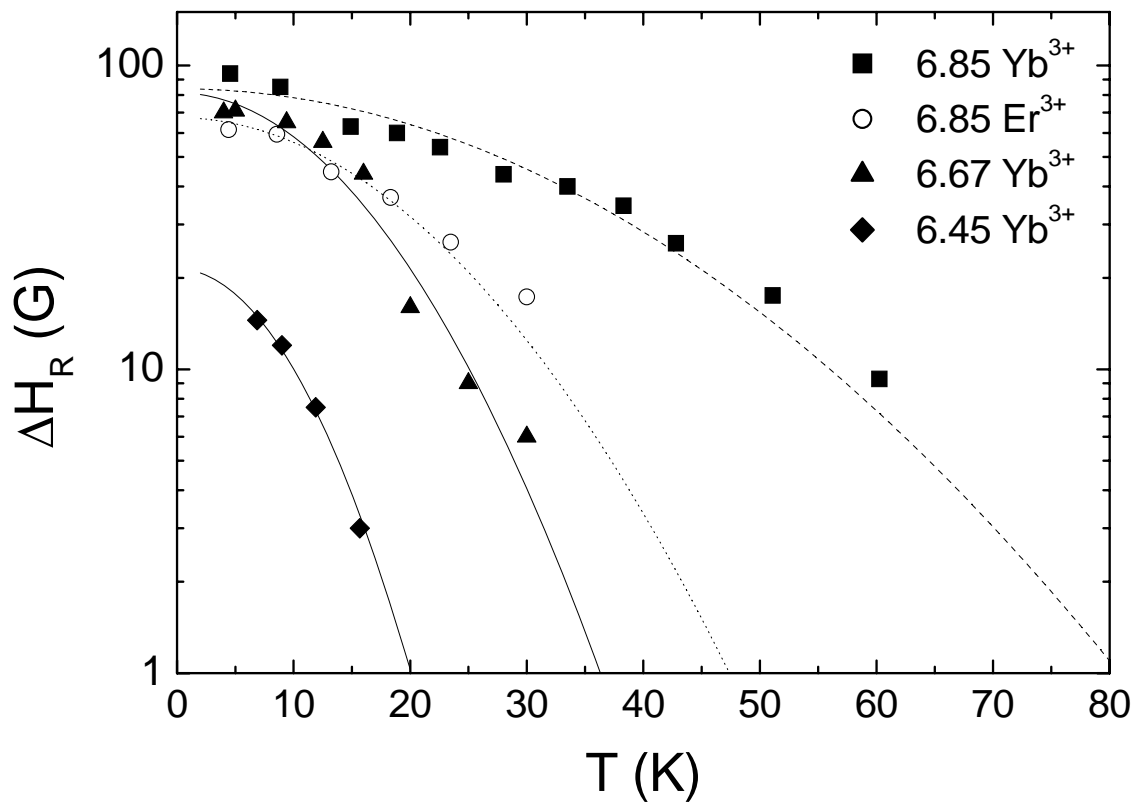

Fig. 3. Temperature dependencies of $\Delta H_R = H_R^{up} - H_R^{down}$. The lines are correspond to equation (5).

Figure 4

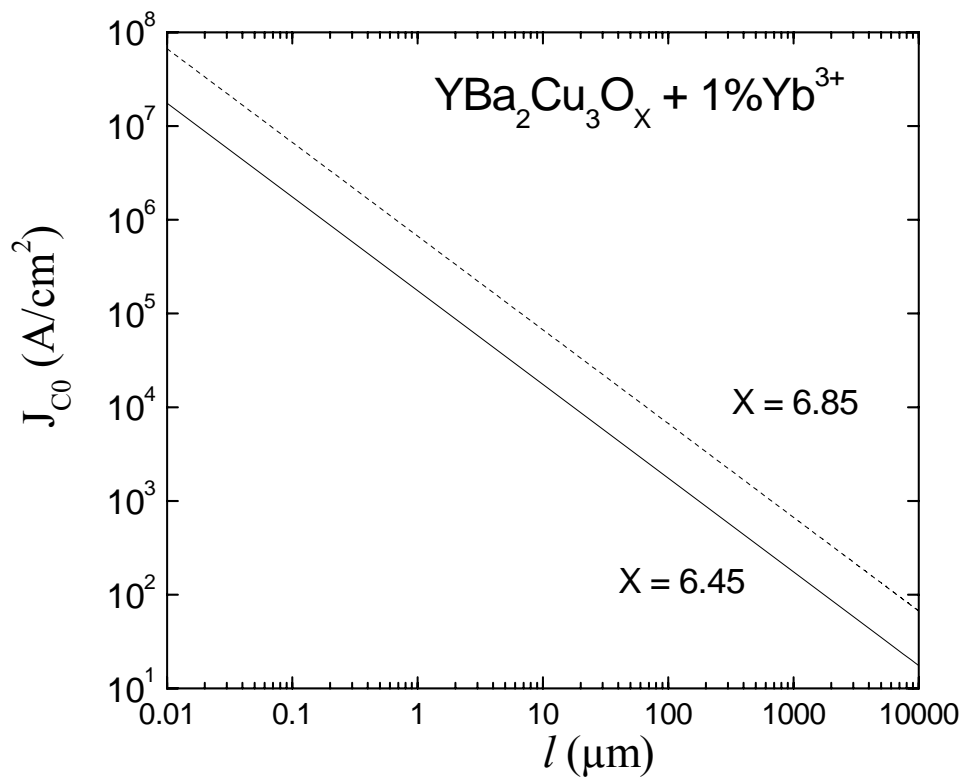

Fig. 4. The relation between the critical current density at T ≈ 5 K ($J_{C0}$) and characteristic spatial scale of the inhomogeneity ($l$) as derived from Eq. (4) for two superconductive samples with different oxygen contents.

Figure 5

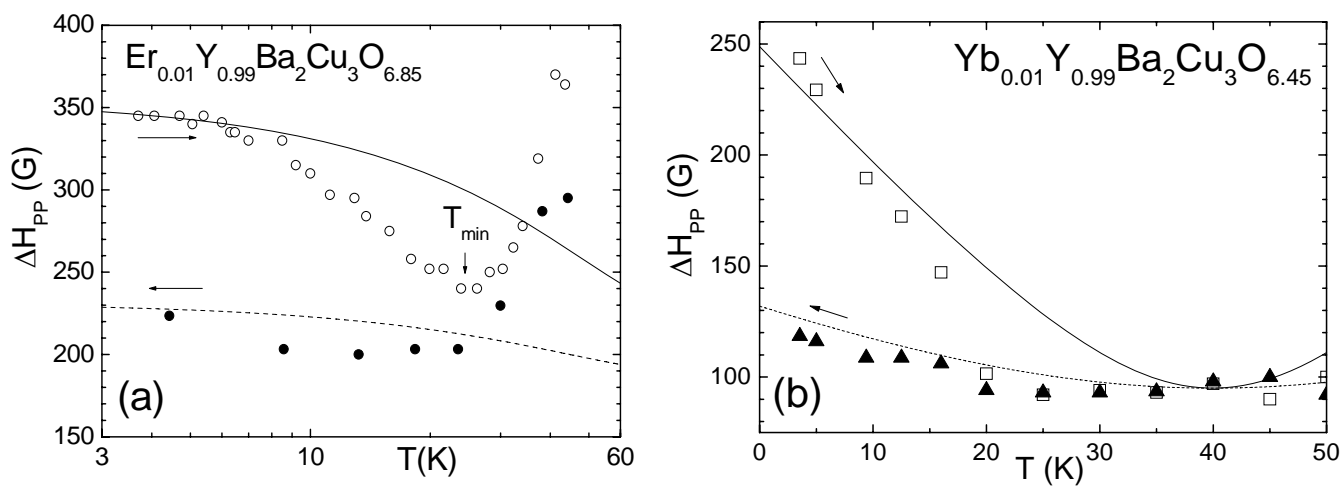

Fig. 5. The temperature dependencies of the linewidth for two investigated samples. The solid lines are correspond to Eq. (9) and (11) with $\lambda_0 = 0.11$ μm, dashed lines - to the same equations with $\lambda_0 = 0.15$ μm (a) and 0.18 μm (b).